# MATHEMATICAL PRINCIPLES OF PREDICTING THE PROBABILITIES OF LARGE EARTHQUAKES

V.M.Ghertzik

*International Institute of Earthquake Prediction*

*Theory and Mathematical Geophysics, Russian Academy of Sciences*

## ABSTRACT

A multicomponent random process used as a model for the problem of space-time earthquake prediction; this allows us to develop consistent estimation for conditional probabilities of large earthquakes if the values of the predictor characterizing the seismicity prehistory are known. We introduce tools for assessing prediction efficiency, including a separate determination of efficiency for «time prediction» and «location prediction»: a generalized correlation coefficient and the density of information gain. We suggest a technique for testing the predictor to decide whether the hypothesis of no prediction can be rejected.

## 1. INTRODUCTION

The probabilistic approach to seismicity presented in this paper requires a certain specification of terminology. The term "earthquake prediction" is not quite correct. In the strict sense, it means the prediction of the place and the time of a forthcoming large earthquake with a certain accuracy. Such prediction, however, is no more possible than prediction of the side on which a coin will fall down at the next toss. Although the laws governing seismicity are obviously deterministic, the sequence of earthquakes looks random to us. This takes place because the information on the processes in the seismogenic medium and the history of its states are inevitably incomplete. In addition the mechanism of the seismic process contains possibly some stochasticity generators, as can be supposed due to the presence of dynamic chaos in the spring-block models of seismicity (e.g., see [Bak et al., 1989; Carlson and Langer, 1989; Feder and Feder, 1991;

Gabrielov et al., 1994]) and, in particular, of the strange attractor found in the spring-block model with healing [Gertzik, 2003].

As a result, it is possible to predict only earthquake probabilities but not earthquakes themselves [Geller, 1997]. One of the goals of this work is to estimate the local conditional probability of a large earthquake where the condition is the observable prehistory of physical processes in the medium.

Practically all of the really applied prediction algorithms use the seismic pre-history in a "convolved" form, which is a function of time and coordinates represented as a 1-D functional of a catalog of preceding earthquakes. In this paper, such functions are called predictors. Examples of such functionals can be found in [Kagan and Jackson, 2000; Kossobokov et al., 1999; *Nonlinear Dynamics ...,* 2003; Sobolev et al., 1996, 1999; Harte et al., 2003]. These functionals take either real values [Kagan and Jackson, 2000; Sobolev et al., 1996, 1999; Harte et al., 2003] or, in the degenerate case, the values 1 (alarm) or 0 (no alarm) [Kossobokov et al., 1999; *Nonlinear Dynamics ...,* 2003].

None of the known algorithms solves the problem of estimating earthquake probabilities from predictors, although some of the algorithms [Kagan and Jackson, 2000; Sobolev et al., 1996, 1999; Harte et al., 2003] formally calculate the probabilities. However, the latter probabilities are not estimates of the true probabilities, but model constructions, in other words, predictors normalized as probabilities.

The problem of earthquake prediction, obviously, divides into two almost independent problems. The first consists in the construction of a predictor which has the highest «correlation» with a future large earthquake. The quality of its solution depends on our understanding of the physical nature of earthquake generation. The second problem is to develop mathematical tools suitable for estimating the conditional probabilities of large earthquakes with known predictor values and to compare the quality of various predictors. This paper is devoted to the solution of the second problem.

The difficulties involved in the estimation of the conditional probabilities of large earthquakes are associated primarily with the absence of an adequate mathematical model. The model of seismicity as a marked random point process reflecting an earthquake catalog is successfully developed by Vere-Jones [1995, 1998]. An important step is his proposal to use information gain as a characteristic of the prediction efficiency for one-dimensional case. However, the point process appears to be an object that is not well suited for estimation of conditional probabilities.

The problem is substantially simplified by representing the predictor as a stationary ergodic multicomponent random process. The components of the latter are interacting random functions of discrete time, defined at the gridpoints of a spatial lattice. The values of functions are functionals of preceding physical processes, not necessarily represented in the same form. This representation enables us to solve the following problems that could not be solved previously:

(1) consistent estimation of conditional probabilities of large earthquakes at a earthquakes given value of the predictor; the estimation of stationary probabilities of large earthquakes (seismic hazard);

(2) development of tools for determining the efficiency of predictors, namely, the generalized correlation coefficient and information gain density, each enabling an unambiguous selection of the most preferable algorithm; the possibility to determine separately the efficiency of time prediction and location prediction;

(3) development of a predictor testing method in order to determine whether the hypothesis of the absence of prediction can be rejected (otherwise, the predictor is not considered).

These results are set forth in this paper.

Note that the introduction of the generalized correlation factor allows us to use correlation methods and factor analysis in application to precursors, to rank them and to reject noninformative precursors.

Any of the known prediction algorithms contains uncontrollable error. The method proposed here for calculating of for calculating probabilities of large earthquakes is not an exception; however, if the stationarity and ergodicity of the process hold, this error tends to zero as time goes on.

Strict mathematical formulations can create the impression that some new limiting requirements are made to apply to the seismic process. This impression would be false: the conditions imposed here are implicitly used in all existing prediction procedures.

The following generally accepted symbols are used here: $\mathrm{Pr}\{A\}$ is the probability of a set $A$, and $\mathrm{Pr}\{A|B\}$, $\mathrm{Pr}\{A|y\}$ are the conditional probabilities of $A$ given the set $B$ or the random value $y$.

**Remark 1**. The theory of Markovian processes with interactions on a spatial lattice arose as a branch of probability theory in the late 1960s. Glauber [1963] was the first to introduce these processes in a physical analysis, and their mathematical description was initiated by Stavskaya and Pyatetskii-Shapiro [1968], Spitzer [1969], and Dobrushin [1971a, 1971b]. Since then the theory quickly grew and developed, finding unexpected links with other fields of science. The initial stimulus of its development was statistical mechanics. Eventually it became clear that models of a very similar mathematical structure can naturally arise in other contexts, such as neural networks, tumor growth, the spread of infections, behavioral systems, etc. Now it is possible to conclude that multicomponent random processes are also a useful model in the theory of earthquake prediction.

## 2. PREDICTION PROBLEM

Let a region on the Earth's surface be divided into identical squares $s_\mathbf{x}$ of area $s$ with centers $\mathbf{x}$ belonging to the subset $A$ of an integral-value lattice with a step $a$, $\mathbf{x} \in A$. (Details related to the Earth's surface curvature are omitted here to make the description less cumbersome.) The time interval $[0,T]$ on which the seismic process is considered is divided into nonintersecting segments $[t_k, t_{k+1})$, $t_{k+1} = t_k + \Delta t$, $k = 0,...,K$, $t_0 = 0$, $t_N = T - \Delta t$.

The prediction algorithms, whose construction problems will be discussed in another paper, specify the rules of calculating the values of the function $f \equiv f(\mathbf{x}, t_k)$ of the coordinates $(\mathbf{x}, t_k)$, $\mathbf{x} \in A$, $k = 0,...,K$; $f$ is the functional of seismic prehistory up to the time $t_k$. We call this function a *predictor*. (If the voting procedure of criteria is used, such function is the voting result accepting two values, 0 and 1)

We introduce an *indicator of events*: the two-value function $h \equiv h(\mathbf{x}, t_k)$, $\mathbf{x} \in A$, $k = 0,...,K-1$, defined by the following rule: $h(\mathbf{x}, t_k) = 1$, if an *event* occurred in the square $s_\mathbf{x}$ in the time interval $[t_k, t_{k+1})$, i.e., by definition , at least one large earthquake with a magnitude of $M \geq M_0$, and $h(\mathbf{x}, t_k) = 0$ otherwise. It is important to emphasize that $f(\mathbf{x}, t_k)$ is a characteristic of the history of geophysical processes in the earth till the moment $t_k$ and the function $h(\mathbf{x}, t_k)$ describes the events taking place in the future in relation to $t_k$. (The definition of "large" earthquake is not essential for most purposes of this work; instead, one can use an indicator of any rare event that depends on the prehistory, extending this approach to other types of time-space prediction. Only the testing procedure has the formal requirement for an approximately Poissonian property of the sequence of large earthquakes.)

The prediction problem can now be formulated as follows. Using the information contained in the functions $f(\mathbf{x}, t_k)$, $h(\mathbf{x}, t_k)$, $\mathbf{x} \in A$, $k = 0,...,K-1$ we have to construct estimates of estimates of the conditional probabilities $\Pr\{h(\mathbf{x}, t_K) | f(\mathbf{x}', t_k)$, $\mathbf{x}' \in A$, $k = 0,...,K-1\}$, $\mathbf{x} \in A$ that at least one large earthquake will occur in the square $s_\mathbf{x}$ during the period $[t_K, t_{K+1})$ given the values of the predictor at preceding times. In this formulation, the prediction problem is obviously recurrent with respect to $K$.

Under the assumption of stationarity and ergodicity of the seismic process, the prediction problem also includes the calculation of current estimates of the stationary (unconditional) probabilities of events, i.e., the estimation of seismic hazard. (By stationarity, as usual, we mean the stationarity in time but not in space. The spatial "stationarity" is referred to as invariance under translations.) Note that, while the unconditional probabilities themselves are assumed to be constant, their estimates vary with time, approaching the true values asymptotically with increasing $N$.

**Remark 2**. The squares $s_\mathbf{x}$ and, more generally, spatial domains in which the probabilities are estimated must not intersect; otherwise, two different estimates will be obtained in the intersection region and no authentic rule exists for discriminating between them. One can assume, for example, that the vertical square boundaries belong to the left squares and the horizontal boundaries belong to the lower squares. The parameters of time and space discretization $a$ and $\Delta t$ can be chosen in two ways. First, they can be preset by the researcher because the problem of estimating the earthquake probability during the period $\Delta t$ in a square with the side $a$ makes sense. Second, these parameters can be fitted to maximize the chosen characteristic of the prediction quality for a given predictor (see Sections 5 and 6). Such values of the space-time lattice parameters can be regarded as natural for each "predictor - prediction quality characteristic" pair. Apparently, universally "correct" lattice parameters do not exist.

## 3. MULTICOMPONENT RANDOM PROCESS

To carry out the formulated program as outlined above, it is necessary to formulate a theoretical model of the process for which functions $f$ and $h$ are observable realizations. We define a two-dimensional multicomponent random process $\{\xi, \eta\} \equiv \{\xi_\mathbf{x}, \eta_\mathbf{x}\}$ with the components $\{\xi_\mathbf{x}, \eta_\mathbf{x}\} \equiv \{\xi_{\mathbf{x},k}, \eta_{\mathbf{x},k}\}$ at points $\mathbf{x}$ of the set $A$, $\mathbf{x} \in A$, which are usual random processes with discrete time $k$, $-\infty < k < \infty$. The stochastic functions $\xi_\mathbf{x}$ take the real values, $\xi_{\mathbf{x},k} \in \mathbf{R}^1$ (here and further $\mathbf{R}^1$ is a real straight line), and $\eta_{\mathbf{x}}$ take the values 0 and 1. The values of the predictor $f(\mathbf{x},t_k)$ and the indicator of events $h(\mathbf{x},t_k)$ are considered here as realizations of the processes $\xi_{\mathbf{x},k}$ and $\eta_{\mathbf{x},k}$, respectively. The process $\{\xi,\eta\}$ is assumed to be stationary and ergodic (see Remark 3.) and the following two natural postulates on properties of conditional distributions are introduced into consideration.

***Locality.*** The property of locality means that the conditional probabilities $\mathrm{Pr}_{\mathbf{x}}\{\eta_{\mathbf{x},0}|\xi_{\mathbf{x}',k}, \mathbf{x}' \in A; k \leq 0\}$ of value $\eta_{\mathbf{x},0}$ at a point $\mathbf{x}$ under the condition $\{\xi_{\mathbf{x}',k}, \mathbf{x}' \in A; k \leq 0\}$ depend only on value $\xi_{\mathbf{x},0}$ at the point (the index $\mathbf{x}$ in the notation of probability $\mathrm{Pr}_{\mathbf{x}}$ specifies its possible dependence on coordinates):

$$\mathrm{Pr}_{\mathbf{x}}\{\eta_{\mathbf{x},0}|\xi_{\mathbf{x}',k}, \mathbf{x}' \in A; k \leq 0\} = \mathrm{Pr}_{\mathbf{x}}\{\eta_{\mathbf{x},0}|\xi_{\mathbf{x},0}\}.$$

This equality is a special case of the Markovian property. It expresses the assumption the implicitly used in all known prediction procedures, that the probability of an event (or a rule for issuing an an alarm) depends on the predictor value $\xi_{\mathbf{x},0}$ only. In other words, it is supposed that the predictor accumulates all necessary information on the past on which probabilities of values of the indicator $\eta_{\mathbf{x},0}$ depend. This is a natural assumption because the predictor is usually constructed just to take this information into account. Since the predictor actually contains only part of the necessary information, the prediction quality depends on how large this part is and, as is shown below, can be estimated by it.

Due to stationarity, the time variable $k$ is omitted in $\xi_{\mathbf{x},k}$ and $\eta_{\mathbf{x},k}$. A stationary multicomponent random process considered as a multidimensional random variable with spatial components is usually referred to as a random field, in this case defined on a finite subset of a 2-dimensional lattice.

***Conditional translational invarianc*** rule for issuing an an alarm) depends on the predictor value $\xi_{\mathbf{x},0}$ only. In other words, it is**.** Below we assume that the conditional probability $\mathrm{Pr}_{\mathbf{x}}\{\eta_{\mathbf{x}}|\xi_{\mathbf{x}}\}$ is invariant under spatial translations, i.e., depends only on $\eta_{\mathbf{x}}$ and $\xi_{\mathbf{x}}$ and does not depend on $\mathbf{x}$:

$$\mathrm{Pr}_{\mathbf{x}}\{\eta_{\mathbf{x}}|\xi_{\mathbf{x}}\} \equiv \mathrm{Pr}\{\eta_{\mathbf{x}}|\xi_{\mathbf{x}}\}.$$

If the condition is not to use predictor values, but all of prehistory, i.e. the full set of functionals that define the seismic process (including local properties of the earth), then conditional translational invariance is a simple consequence of a fundamental methodological principle of scientific research, according to which identical causes (under identical conditions which, however, can be included formally in the causes) produce identical effects irrespective of the place and time of their occurrence.

Conditional translational invariance is the only assumption which can be made without using additional nontrivial information on hidden parameters (any form of dependence of the conditional distribution on the coordinates is just such information). It is important to note that it is not a "law of nature" but only an idealization caused by lack of the information and it is not necessarily true if the lack is filled. In the latter case the region should be divided into smaller areas for which the assumption of conditional translational invariance remains valid. Then the problem should be considered separately for each area. Note, however, that the property of conditional translational invariance applies only to conditional probabilities and is in no way related to the spatial seismicity heterogeneity. The latter is expressed by *unconditional* probabilities of the predictor $\Pr_{\mathbf{x}}\{\xi_{\mathbf{x}}\}$ and by unconditional probabilities of the events $\Pr_{\mathbf{x}}\{\eta_{\mathbf{x}}\}$, which remain spatially heterogeneous. Furthermore, this property is implicitly used in all known prediction techniques because the probability of an event (or the rule for issuing an alarm) is determined only by the predictor value $\xi_{\mathbf{x},0}$ and does not depend on the position of the point $\mathbf{x}$ in the prediction region.

A complete description of the process requires knowledge of all its finite-dimensional distributions. For purposes of prediction, however, it is sufficient to consider the stationary probabilities $\Pr_{\mathbf{x}}\{\xi_{\mathbf{x}} \le u; \eta_{\mathbf{x}} = \theta\}$ that $\xi_{\mathbf{x}} \le u$ and $\eta_{\mathbf{x}} = \theta$, i.e. marginal distributions of the components of the multidimensional stochastic function (a random field). The conditional probability $\Pr_{\mathbf{x}}\{\eta_{\mathbf{x}} = 1 \mid \xi_{\mathbf{x}} = u\}$ that at least one earthquake with magnitude $M \ge M_0$ will occur in the square $s_{\mathbf{x}}$ in the time interval $\Delta t$ under the condition that the predictor takes the value $u$ is denoted as

$$g(u) = \Pr\{\eta_{\mathbf{x}} = 1 \mid \xi_{\mathbf{x}} = u\}. \tag{1}$$

The following notation is used for local distributions:

$$P_{\mathbf{x}}(u) = \text{Pr}_{\mathbf{x}}\{\xi_{\mathbf{x}} \leq u\}, \tag{2}$$

$$P_{\mathbf{x}}(du) = \text{Pr}_{\mathbf{x}}\{u < \xi_{\mathbf{x}} \leq u + du\}. \tag{3}$$

$$\text{Pr}_{\mathbf{x}}\{du, j\} = \text{Pr}_{\mathbf{x}}\{u < \xi_{\mathbf{x}} \leq u + du; \eta_{\mathbf{x}} = j\} \tag{4}$$

Formulas (1) - (3) and the definition of conditional probabilities imply that

$$\text{Pr}_{\mathbf{x}}\{du, 1\} = g(u)\, P_{\mathbf{x}}(du), \tag{5}$$

$$\text{Pr}_{\mathbf{x}}\{du, 0\} = [1 - g(u)]\, P_{\mathbf{x}}(du). \tag{6}$$

The unconditional probability $Pr_{\mathbf{x}}\{\eta_{\mathbf{x}} = 1\}$ of an event in the square $s_{\mathbf{x}}$ is equal to

$$\pi_{\mathbf{x}} \equiv \text{Pr}_{\mathbf{x}}\{\eta_{\mathbf{x}} = 1\} = \int\limits_{u \in \mathbf{R}^1} g(u)\, P_{\mathbf{x}}(du). \tag{7}$$

If a predictor is absent, which is equivalent to its identical equality to a constant, then $g$ also is constant and, as follows from (7), $\pi_{\mathbf{x}} \equiv g$ does not depend on $\mathbf{x}$ and is equal to its spatial average

$$\Pi = \frac{1}{|A|} \sum_{\mathbf{x} \in A} \pi_{\mathbf{x}} = g. \tag{8}$$

The estimate for $\Pi$ is the ratio of the number of events to the number of space-time cells in a given time interval and a given area of the region.

**Remark 3.** The observed plate tectonics does not suppose stationarity and ergodicity of geophysical processes even for geometric reasons (plate motions cannot always maintain their directions without fracture, i.e., without plate changes). These processes can only be considered as quasi-stationary; i.e., one may suppose that there exists a stationary and ergodic process virtually coinciding with the real process over a long time interval. Rejection of this assumption calls into question the very possibility of prediction, which is based on the fact that the earthquake probability is calculated in the future by the same method as in the past and is therefore stationary. Ergodicity is necessary, first, to prove the consistency of an estimate, otherwise one cannot state that the estimate is at all related to the real probability. Second, probability averages cannot be replaced by time averages without ergodicity; i.e., even such simple prediction characteristics as the relative space-time of alarms or the relative number of failures-to-predict cannot be correctly calculated and it is impossible to judge about the quality of the prediction. Stationarity and ergodicity are thus actually assumed in all prediction research research even if it is not mentioned explicitly. In periods of change in the seismic regime (nonstationarity), the quality of the prediction is certain to be lower because these conditions will no longer hold.

**Remark 4**. If it is necessary to get a rough estimate of the probability of events on any subset of squares $s_\mathbf{x}$, $\mathbf{x} \in B$, $B \subseteq A$, one has to adopt an auxiliary postulate on the *conditional independence of events*: at given values of the predictor $\xi_\mathbf{x}$, $\mathbf{x} \in A$, the random variables $\eta_\mathbf{x}$, $\mathbf{x} \in A$, are independent; i.e., taking into account the first two postulates,

$$\Pr\{\eta_\mathbf{x}, \mathbf{x} \in A | \xi_\mathbf{x}, \mathbf{x} \in A\} = \prod_{\mathbf{x} \in A} \Pr\{\eta_\mathbf{x} | \xi_\mathbf{x}\}.$$

(Note that the values of the predictor $\xi_\mathbf{x}$ depend on each other and this assumption means that all dependence relations within seismicity reduce to the interdependence of these values, and events in the period $\Delta t$, i.e., large earthquakes, depend only on the values of $\xi_\mathbf{x}$ and are independent of each other.) This condition is exact at sufficiently small $\Delta t$ because the rate of information propagation in the earth is finite. For the time intervals actually used in prediction, it is only a first approximation under the assumption that the deviations from independence do not affect too much the results of calculations. In this case the researcher has two options only: to accept this assumption or to give up completely the possibility of calculation of joint probabilities of events. Even for the study of conditional pair correlations $\eta_\mathbf{x}$ and $\eta_{\mathbf{x}'}$ it is necessary to have a significant number of events (earthquakes of a sufficiently high magnitude) in each of the fixed squares $s_\mathbf{x}$ and $s_{\mathbf{x}'}$, while in the time intervals actually considered, this number is most often equal to 0 and exceeds 1 only in exceptional cases. (Although this supposition is an idealization from the theoretical point of view, the space-time coordinates of earthquakes with magnitude more than 6.5 in California are such that none of the known statistical tests for the independence of sample elements allows one to reject the hypothesis of their independence. This fact is an experimental reason for admission of the approximation here suggested.)

## 4. ESTIMATING THE DISTRIBUTIONS OF PREDICTORS AND CONDITIONAL PROBABILITIES OF EVENTS

The estimates $P_{K,\mathbf{x}}(u)$ for 1-D distributions of $P_{\mathbf{x}}(u)$ are ordinary empirical distributions:

$$P_{K,\mathbf{x}}(u) = n_{\mathbf{x}}(u)/K,$$

where $n_{\mathbf{x}}(u)$ is the number of values of $f(\mathbf{x},t_k)$, $k = 0,...,K\text{-}1$ that do not exceed $u$. Due to stationarity and ergodicity of the process, these estimates are consistent, i.e., converge to $P_{\mathbf{x}}(u)$ as $K \to \infty$.

An estimate of the conditional probability $g(u)$ can be obtained in various ways. The method proposed here is unambiguously determined by the structure of available data. As an estimate $g_N(u)$ of the function $g(u)$ we use its step approximation $g_{N,\varepsilon}(u)$ obtained as follows. Let $I$ be the number of points $(\mathbf{x},t_k)$, $\mathbf{x} \in A$, $k = 0,...,K\text{-}1$ where $h(\mathbf{x},t_k) = 1$. The symbols $y_i$, $i = 1,...,I$ denote the values of $f(\mathbf{x},t_k)$ at these points arranged in ascending order. We will consider the numbers $n(y_i)$ of all points $(\mathbf{x},t_k)$, $\mathbf{x} \in A$, $k = 0,...,K\text{-}1$ where $f(\mathbf{x},t_k) < y_i$. For a fixed small number $\varepsilon$ we consider the values $\delta_i$, $i = 1,...,I+1$ given by the ratios $\delta_1 = n(y_1)/K_A$; $\delta_i = [n(y_i) - n(y_{i-1})]/K_A$, $i = 2,...,I$; $\delta_{I+1} = 1 - n(y_I)/K_A$, where $K_A = K|A|$, $|A|$ is the number of points in the set $A$. (These values are the relative amounts of points at which $f$ lies between the corresponding values $y_i$.) The subsequent procedure is defined by the following algorithm:

1. The values of $\delta_i$ are inspected in the order of increasing $i$. If $\delta_{i*} < \varepsilon$, then $y_{i*}$ is removed from the sequence of $y_i$, $i = 1,...,I$; if $\varepsilon \le \delta_I$ and $\delta_{I+1} < \varepsilon$, then $y_I$ is removed.

2. The remaining $J$ numbers of $y_i$ are used to construct a new nondecreasing sequence $z_j$, $j = 1,...,J$, $(J \le I)$ and the corresponding sequence $\delta_j$, $j = 1,...,J+1$.

3. If $\varepsilon \le \delta_j$ for all $j = 1,...,J+1$, the procedure is carried out; otherwise, after replacing $z$ by $y$ and the indices $j, J$ by $i, I$ we go to step 1.

(The point of the construction is to eliminate intervals with "weights" smaller than $\varepsilon$. If this is not done, the weights of the intervals will decrease indefinitely, as their number increases with time, and it is unclear how to demonstrate the consistency of the estimates.)

Let $M$ be the number of points $(\mathbf{x}, t_k)$, $\mathbf{x} \in A$, $k = 0, ..., K-1$, at which $h(\mathbf{x}, t_k) = 1$, and let $m(u)$ be the number of those for which $f(\mathbf{x}, t_k) < u$. We define $g_j$, $j = 1, ..., J+1$ by the expressions

$$g_1 = \begin{cases} [1 + m(z_1)] / n(z_1) & \text{for} \quad n(z_1) > 0; ; \\ 1 & \text{for} \quad n(z_1) = 0. \end{cases}$$

$$g_j = [m(z_j) - m(z_{j-1})] / [n(z_j) - n(z_{j-1})], j = 2, ..., J;$$

$$g_{J+1} = [M - m(z_J)] / [K_A - n(z_J)].$$

The desired estimate $g_{K,\varepsilon}(u)$ is given by

$$g_{K,\varepsilon}(u) = \begin{cases} g_1 & \text{for} \quad u \leq z_1; \\ g_j & \text{for} \quad z_{j-1} < u \leq z_j, \quad j = 2, ..., J; \\ g_{J+1} & \text{for} \quad z_J < u. \end{cases}$$

(Note that all values $g_j$ are strictly positive by construction. We preferred to avoid zero values because zero is the lower bound of the confidence interval for a positive probability at any estimation accuracy. Here the probability is naturally estimated by the corresponding rates.)

Using the ergodicity of the process and the natural restraints on the form of $g(u)$ и $P_\mathbf{x}(u)$ (the simplest of them is that $g(u)$ vanishes outside a bounded segment of the axis $u$ and is strictly positive and uniformly continuous on it; also, that the distributions of $P_\mathbf{x}(u)$ have densities), one can show that $g_{K,\varepsilon}(u)$ converges to the step function $g_\varepsilon(u)$ as $K \to \infty$ for all u and $g_\varepsilon(u)$ uniformly converges to $g(u)$ as $\varepsilon \to 0$. In this sense the estimate $g_{K,\varepsilon}(u)$ is consistent.

Below, to make the notation shorter, we use expressions containing only theoretical probabilities (1)--(4). To obtain empirical estimates of these expressions, the above probability estimators $P_{\mathbf{x}}(u)$ and $g(u)$ should be substituted into the formulas.

## 5. GENERALIZED CORRELATION COEFFICIENT AS A CHARACTERISTIC OF PREDICTION EFFICIENCY

In the one-component (1-D) case, when $|A| = 1$ and the spatial indices of $\mathbf{x}$ can be omitted, the prediction efficiency of the predictor $\xi_{\mathbf{x}}$ can be expressed by the correlation coefficient $\rho(g, \eta)$ between the conditional probability $g(u) = \Pr\{\eta = 1 | \xi = u\}$ that the event will occur given that the predictor is equal to $u$ and the indicator of events $\eta$. It is easy to show that $\rho(g, \eta) = 0$ in the case of independent $\xi$ and $\eta$ and $g(\xi) = \eta$, $\rho(g, \eta) = 1$ in the case of (impossible) error-free prediction.

In order to extend the concept of the correlation coefficient to the multicomponent case, we specify the field of mathematical expectation for an arbitrary function $\varphi(\xi_{\mathbf{x}}, \eta_{\mathbf{x}})$:

$$E_{\mathbf{x}} \varphi(\xi_{\mathbf{x}}, \eta_{\mathbf{x}}) = \sum_{j=0,1} \int \varphi(u, j) \Pr_{\mathbf{x}}\{du, j\}, \qquad (9)$$

and the space-averaged mathematical expectation

$$<\varphi(\xi, \eta)>_A = \frac{1}{|A|} \sum_{\mathbf{x} \in A} E_{\mathbf{x}} \varphi(\xi, \eta_{\mathbf{x}}) \qquad (10)$$

We introduce the random fields $\eta_{\mathbf{x}}^* = \eta_{\mathbf{x}} - <\eta>_A$ and $g^*(\xi_{\mathbf{x}}) = g(\xi_{\mathbf{x}}) - <g(\xi)>_A$, $\mathbf{x} \in A$, and define a generalized correlation coefficient $\rho_A(\eta, g)$ of the random fields $\eta_{\mathbf{x}}$ and $g(\xi_{\mathbf{x}})$ by the expression

$$\rho_A(\eta, g) = \frac{<\eta^* g^*(\xi)>_A}{\sqrt{<\eta^{*2}>_A <g^{*2}(\xi)>_A}}. \qquad (11)$$

Using the equalities $\eta_{\mathbf{x}} = \eta_{\mathbf{x}}^2$, $E_{\mathbf{x}} \eta_{\mathbf{x}} = \Pr_{\mathbf{x}}\{\eta_{\mathbf{x}} = 1\}$ and expression (7), we find

$$E_{\mathbf{x}} \eta_{\mathbf{x}} = E_{\mathbf{x}} \eta_{\mathbf{x}}^2 = \Pr_{\mathbf{x}}\{\eta_{\mathbf{x}} = 1\} = \int \Pr_{\mathbf{x}}\{du, 1\} = \int g(u) P_{\mathbf{x}}(du),$$

which, taking into account (8), (9), and (10), implies

$$<\eta>_A = <\eta^2>_A = <g(\xi)>_A = \frac{1}{|A|} \sum_{\mathbf{x} \in A} \int g(u) P_{\mathbf{x}}(du) = \Pi . \tag{12}$$

The equalities following from (5)

$$E_{\mathbf{x}} g \eta_{\mathbf{x}} = \int g(u) \mathrm{Pr}_{\mathbf{x}}\{du, 1\} = \int g^2(u) \mathrm{Pr}_{\mathbf{x}}\{du\}$$

give with the help of (9) и (10)

$$<g^2(\xi)>_A = <g(\xi)\eta>_A = \frac{1}{|A|} \sum_{\mathbf{x} \in A} \int g^2(u) P_{\mathbf{x}}(du). \tag{13}$$

Using the definitions of $g^*$ и $\eta^*$ and (12) и (13) it is easy to obtain the expressions

$$<\eta^{2*}>_A = \Pi(1-\Pi), \tag{14}$$

$$<g^*(\xi)\eta^*>_A = <g^{*2}(\xi)>_A = \frac{1}{|A|} \sum_{\mathbf{x} \in A} \int g^2(u) P_{\mathbf{x}}(du) - \Pi^2.$$

Substituting them into (11), we obtain the formula for the generalized correlation coefficient

$$\rho_A(g, \eta) = \sqrt{\frac{\frac{1}{|A|} \sum_{\mathbf{x} \in A} \int [g(u) - \Pi]^2 P_{\mathbf{x}}(du)}{\Pi(1-\Pi)}}. \tag{15}$$

The denominator here does not vanish because large earthquakes are rare and the real values of $\Pi$ are much smaller than 1, and $\Pi$ is greater than 0 because of the hypothesis that large earthquakes exist in the prediction region.

If the predictor $\xi$ and the event indicator $\eta$ are independent, the conditional probability $g(\xi_{\mathbf{x}})$ does not depend on $\xi_{\mathbf{x}}$, $g(\xi_{\mathbf{x}}) \equiv g_0$, $\mathbf{x} \in A$. It follows from (7) and (11) that $g_0 \equiv \Pi$ and $\rho_{\mathbf{A}}(g, \eta) = 0$. On the other hand, the equality $\rho_{\mathbf{A}}(g, \eta) = 0$ and formula (15) imply that $g(\xi_{\mathbf{x}}) = \Pi$ holds almost certainly for any $\mathbf{x}$, i.e., that $\xi$ and $\eta$ are independent. In this case, the independence of $\xi$ and $\eta$ and their being uncorrelated are equivalent.

For error-free prediction, the equality $g(\xi_{\mathbf{x}}) = \eta_{\mathbf{x}}$ is true. It expresses the fact that, if an event in the lattice cell $s_{\mathbf{x}}$ occurs ($\eta_{\mathbf{x}} = 1$), error-free prediction must define its conditional probability as $g(\xi_{\mathbf{x}}) = 1$, and, if it does not occur, then $g(\xi_{\mathbf{x}}) = 0$. Therefore, $<g^*(\xi)\eta^*>_A = <\eta^{2*}>_A$, and, as follows from (11), (14) and (15), $\rho_A(g, \eta) = 1$.

The quantity $\rho_A(g,\eta)$ has properties quite similar to those of the ordinary correlation coefficient, and its absolute value indicates how close the prediction given by the predictor is to error-free prediction and how far it is from purely random prediction. Thus, the generalized correlation coefficient is a fairly effective characteristic of prediction quality.

Now, we examine the "prediction of event location" defined by stationary probabilities (7). To do this, we assume that the distribution functions $P_{\mathbf{x}}(u)$ of the predictor $\xi_{\mathbf{x}}$ are concentrated at the points $\pi_{\mathbf{x}}$ and $g(\pi_{\mathbf{x}}){=}\pi_{\mathbf{x}}$. Then

$$\rho_A^s(g,\eta) = \sqrt{\frac{\dfrac{1}{|A|}\sum_{\mathbf{x}\in A}\left[\pi_{\mathbf{x}}-\varPi\right]^2}{\left(1-\varPi\right)\varPi}},$$

i.e., as could be expected, the prediction given by the probabilities $\pi_{\mathbf{x}}$ is nontrivial. The parameter $\rho_A^s(g,\eta)$ characterizes the quality of "seismic regionalization" $\pi_{\mathbf{x}}$, which is in fact the "location prediction". The correlation coefficient $\rho_A^s(g,\eta)$ vanishes only in the case of a trivial predictor equal to a constant or in the case of spatially uniform seismicity, when knowledge of $\pi_{\mathbf{x}}=\varPi$ adds nothing to the a priori "ignorance".

As a characteristic describing the quality of time prediction of events, it is natural to choose the quantity

$$\rho_A^t(g,\eta) = \sqrt{\frac{\dfrac{1}{|\mathrm{A}|}\sum_{\mathbf{x}\in\mathrm{A}}\int[g(u)-\pi_{\mathbf{k}}]^2 P_{\mathbf{x}}(du)}{\varPi(1-\varPi)}}$$

so that

$$\rho_{\mathbf{A}}(g,\eta)^2 = \rho_A^t(g,\eta)^2 + \rho_A^s(g,\eta)^2.$$

The consistent estimators $g_{N,\varepsilon}(\xi)$ converge as $K\to\infty$ и $\varepsilon\to 0$ to the true (a priori unknown) conditional probability $g(\xi_{\mathbf{x}})$, which is controlled by the nature of the probabilistic dependence of events and the predictor only, but is not constructed from the latter. We show that the generalized correlation coefficient $\rho_A(g_1, \eta)$ for the "model" conditional probabilities $g_1(\xi_{\mathbf{x}})$ constructed from the predictor $\xi$ by any method cannot exceed $\rho_A(g_1, \eta)$. In the linear space of the functions $\varphi(\xi, \eta) \equiv \{\varphi(\xi_{\mathbf{x}}, \eta_{\mathbf{x}}), \ \mathbf{x} \in A\}$ we introduce the scalar product $(\varphi, \psi) = <\varphi(\xi, \eta)\psi(\xi, \eta)>_A$ (it is easy to verify that all properties of the scalar product are satisfied). Then by using the notation $\varphi^*(\xi, \eta) = \varphi(\xi, \eta) - <\varphi(\xi, \eta)>_A$ and the easily verifiable equalities $(\eta^*, g^*) = (g^*, g^*)$ и $(\eta^*, g_1^*) = (g^*, g_1^*)$ we have

$$\rho_A(g, \eta) - \rho_A(g_1, \eta) \ = \frac{(\eta^*, g^*)}{\sqrt{(\eta^*, \eta^*) \ (g^*, g^*)}} - \frac{(\eta^*, g_1^*)}{\sqrt{(\eta^*, \eta^*) \ (g_1^*, g_1^*)}} =$$

$$= \frac{(\eta^*, g^*) \ \sqrt{(g_1^*, g_1^*)} - (\eta^*, g_1^*) \ \sqrt{(g^*, g^*)}}{\sqrt{(\eta^*, \eta^*) \ (g_1^*, g_1^*) \ (g^*, g^*)}} = \frac{\sqrt{(g_1^*, g_1^*) \ (g^*, g^*)} - (g^*, g_1^*)}{\sqrt{(\eta^*, \eta^*) \ (g_1^*, g_1^*)}}$$

.

The difference in the numerator of the last ratio is nonnegative as a consequence of the Cauchy-Schwarz inequality. Therefore

$$\rho_A(g, \eta) - \rho_A(g_1, \eta) \geq 0,$$

and the statement is proven.

.

The scalar product defines the norm $\|\varphi\| = (\varphi, \varphi)^{\frac{1}{2}}$ and the metric $r(\varphi, \psi) = \|\varphi - \psi\|$. The event indicator $\eta$ coincides with the conditional probability $g$ in the case of error-free prediction. Therefore the distance $r(g, \eta)$ between the conditional probability $g$ given by the predictor and the $\eta$ is also a characteristic of the predictor quality and is expressed by (12)-(15) in terms of the generalized correlation coefficient as follows:

$$r(g, \eta) \ = <(g(\xi) - \eta)^2>_A^{\frac{1}{2}} = \{\Pi(1 - \Pi)[1 - \rho_A(g, \eta)^2]\}^{\frac{1}{2}}.$$

This characteristic, however, has the following significant shortcoming: if we add to the prediction region an aseismic area consisting of the elements $s_{\mathbf{x}}$ with $\pi_{\mathbf{x}} = 0$, the value of $\Pi$ will diminish because the numerator in (11) is left unchanged and the denominator increases. The presence of $\Pi^{\frac{1}{2}}$ on the right-hand side of the last expression also implies a decrease in $r(g, \eta)$ (we recall that $\Pi \ll 1$), creating the false impression that the a prediction quality increased due to the added area. Characteristics with this property (another example is the relative alarm space-time) are only usable for the comparison of predictors within a fixed region. If predictors are defined in different regions, it is incorrect to use such characteristics for their comparison.

The generalized correlation coefficient is free from this shortcoming. Because $\Pi \to 0$ if the area of the added aseismic region increases, we find from $\Pi = \dfrac{1}{|A|} \sum\limits_{\mathbf{x} \in A} \int g(u) P_{\mathbf{x}}(du)$ and (15) that at small $\Pi$ the following decomposition is true:

$$\rho_A(g, \eta) = \sqrt{\frac{\sum\limits_{\mathbf{x} \in A} \int g(u)^2 P_{\mathbf{x}}(du)}{\sum\limits_{\mathbf{x} \in A} \int g(u) P_{\mathbf{x}}(du)}} + o(\Pi).$$

By definition an aseismic area consists of only such values $\xi$ that $g(u) = 0$ with probability 1; therefore its addition does not change the sums in the last formula. Thus, although the addition of an aseismic area to the initial region changes $\rho$, bringing it in correspondence with the new prediction region, it gives no systematic apparent "improvements" of the prediction efficiency.

## 6. INFORMATION GAIN INDUCED BY A PRECURSOR AS A CHARACTERISTIC OF THE PRECURSOR EFFICIENCY

Information gain as a characteristic of prediction efficiency in the one-component case was introduced by Vere-Jones [1998]. Here this characteristic is extended to the spatial multicomponent case. We remind the reader the essence of the notions of entropy and information using an illustrative, though not fully rigorous, construction from [Prokhorov and Rozanov, 1973].

In a very rough approach the information contained in a text is measured by its length. The smallest length $S$ of a line, which consists of 0 and 1 (binary code) and allows one to count $N$ different objects, satisfies the relation

$$0 \leq S - \log_2 N \leq 1.$$

The value of $S \approx \log_2 N$ characterizes the length of the most economical code combinations allowing such count.

Now, we take an experiment whose result can be one of $N$ incompatible events $A_1,\dots,A_N$ with the probabilities $p_1,\dots,p_N$, accordingly, $p_1 +\dots+ p_N = 1$. The report about the outcome of $n$ independent and identical trials resulting in one of these events can be written as a sequence $(A_{i_1},\dots,A_{i_n})$, $A_{i_k}$ being an event in the $k$-th test. Because the frequency $n_i/n$ event $A_i$ occurring in a large series of trials practically coincides with its probability $p_i$, we may assume that $A_i$ occurs $n_i$ times in the $(A_{i_1},\dots,A_{i_n})$. The number of all such highly probable records is

$$N_n = \frac{n!}{n_1!\dots n_N!}$$

and according to the Stirling formula the length of the most economical code combination for their description is approximately equal to

$$S_n \approx \log_2 N_n \approx - n \sum_{i=1}^{N} p_i \log_2 p_i.$$

The quantity $S_n$ characterizes the uncertainty before a series of $n$ trials: only one of $2^{S_n}$ possibilities can be realized. The measure of uncertainty per trial

$$S = S(p_1,\dots,p_N) = - \sum_{i=1}^{N} p_i \log_2 p_i$$

is called the Shannon entropy of the distribution $p_1,\dots,p_N$ (in physics, entropy is the measure of disorder or the measure of chaos). After one trial the uncertainty of the future decreases by $S = S_n - S_{n-1}$ and this decrease forms the gain in the amount of information $I= S$ as a result of the trial.

The quantity

$$S_A(\eta) = - \varPi \log_2 \varPi - (1-\varPi) \log_2 (1-\varPi) \tag{16}$$

is the entropy of the distribution of the event indicator $\eta$ in the space-time lattice cell if a predictor is absent. The conditional entropy $S_A(\eta \,|u)$ given that the predictor $\xi$ has the value $u$ is

$$S_A(\eta \,|u) = - g(u) \log_2 g(u) - [1- g(u)] \log_2 [1- g(u)].$$

We average the conditional entropy over the distributions of the condition (i.e., the predictor) and over space and find the entropy $S_{A,\xi}(\eta)$ of the distribution of the event indicator at known values of the predictor:

$$S_{A,\xi}(\eta) = -\frac{1}{|A|} \sum_{\mathbf{x}\in A} [ \int_{u\in\mathbf{R}^1} g(u) \log_2 g(u)\, P_{\mathbf{x}}(du) + \int_{u\in\mathbf{R}^1} [1 - g(u)] \log_2 [1 - g(u)]\, P_{\mathbf{x}}(du)]. \tag{17}$$

Thus, knowledge of predictor values decreases the uncertainty of the future by the amount $S_A(\eta) - S_{A,\xi}(\eta)$, which is the information gain $I_A(g,\eta)$ provided by the predictor. From (12), (16) and (17) we can see that this increase is equal to

$$I_A(g,\eta) = \frac{1}{|A|} \sum_{\mathbf{x}\in A} [ \int_{u\in\mathbf{R}^1} g(u) \log_2 \frac{g(u)}{\varPi}\, P_{\mathbf{x}}(du) + \int_{u\in\mathbf{R}^1} [1 - g(u)] \log_2 \frac{1 - g(u)}{1 - \varPi}\, P_{\mathbf{x}}(du)].$$

By analogy with the one-component case [Kolmogorov, 1965], it is natural to refer to the quantity $I_A(g,\eta)$ as the information on the random field $\eta$ contained in the random field $\xi$. As in the one-component case, using the Jensen inequality for convex functions [Feller, 1967] and the fact that the function $\frac{1}{|A|} \sum_{\mathbf{x}\in A} P_{\mathbf{x}}(u)$ has the properties of a distribution function, one can easily show that the information $I_A(g,\eta)$ is non-negative and equal to 0 if and only if the random fields $\xi$ and $\eta$ are independent. The information $I_A(g,\eta)$ reaches the maximum value equal to $S_A(\eta)$ in the case of exact prediction.

By construction, $I_A(g,\eta)$ is the specific information per one space-time lattice cell and, therefore, depends on the dimensions of this cell. In order to compare the predictors constructed for cells of different dimensions we introduce the space-time information density

$$i_A(g,\eta) = \frac{1}{s\Delta.t} I_A(g,\eta),$$

which is regarded here as a characteristic of the predictor quality.

The information density $i_A(g,\eta)$ can be represented as the sum

$$i_A(g,\eta) = i_A^s(g,\eta) + i_A^t(g,\eta)$$

of the characteristic of "location prediction"

$$i_A^s(g,\eta) = \frac{1}{|A| s \Delta t} \sum_{\mathbf{x} \in A} \left[ \pi_{\mathbf{x}} \log_2 \frac{\pi_{\mathbf{x}}}{\Pi} + (1 - \pi_{\mathbf{x}}) \log_2 \frac{1 - \pi_{\mathbf{x}}}{1 - \Pi} \right]$$

and the characteristic of "time prediction"

$$i_A^t(g,\eta) = \frac{1}{|A| s \Delta t} \sum_{\mathbf{x} \in A} \left[ \int_{u \in \mathbf{R}^1} g(u) \log_2 \frac{g(u)}{\pi_{\mathbf{x}}} P_{\mathbf{x}}(du) + \int_{u \in \mathbf{R}^1} [1 - g(u)] \log_2 \frac{1 - g(u)}{1 - \pi_{\mathbf{x}}} P_{\mathbf{x}}(du) \right].$$

## 7. TESTING A PREDICTOR

Quality characteristics of a predictor can be calculated with a given accuracy only in the limit of an infinite observation time. For finite time one can only obtain their statistical estimates, and even the construction of confidence intervals is an unsolved problem here. Nonzero values of these estimates do not guarantee nonzero values of the estimated characteristics. The deviation from zero can be a consequence of randomness even if the given predictor does not predict earthquakes at all. Therefore it is necessary to know whether the observed data allow us to reject the hypothesis that the event indicator is independent of the predictor at an acceptable level of significance. Then we shall be able to state that the prediction exists actually with a given probability.

We introduce the formal distributions

$$P^*(u) = \frac{1}{|A|} \sum_{\mathbf{x} \in A} P_{\mathbf{x}}(u)$$

and

$$P'(u) = \frac{1}{|A|} \sum_{\mathbf{x} \in A} \Pr_{\mathbf{x}} \{ \xi_{\mathbf{x}} \le u | \eta_{\mathbf{x}} = 1 \}.$$

The function $F^*(y) = P^*(P^{*-1}(y)) = y$ of the variable $y = P^*(u)$ is the uniform distribution $F^*(y) = \Pr\{\zeta \le y\}$ of a random variable $\zeta$ in the interval $[0,1]$. The function $F(y) =$ $=P'(P^{*-1}(y))$ is a distribution function in [0,1], and in parametric form its points have the abscissas $P^*(u)$ and the ordinates $P'(u)$. If the random fields $\xi$ and $\eta$ are independent, then $P'(u)=P^*(u)$ and the distribution $F(y)$ is uniform. In order to reject the no-prediction hypothesis, i.e., the hypothesis of the independence of $\xi$ and $\eta$ it is sufficient to reject the hypothesis $H_0$ that of the distribution $F(y)$ is uniform.

An empirical distribution corresponding to $F(y)$ can be constructed as follows. In the initial catalog, we remove large (with magnitudes $M \geq M_0$ that are lower than the magnitude of the main shock) foreshocks and aftershocks of the main shocks (with magnitudes higher than $M_0$); in other words, we "stick" them together with the main shocks. Using the notation of Section 4, let $N$ be the number of points $(\mathbf{x},t_k)$, $\mathbf{x} \in A$, $k = 0,...,K\text{-}1$ at which $h(\mathbf{x},t_k) = 1$. Let the symbols $u_n$, $n = 1,...,N$ denote the increasing values of the function $f(\mathbf{x},t_k)$ at these points; we introduce the corresponding numbers $m(u_n)$ of all points $(\mathbf{x},t_k)$, $\mathbf{x} \in A$, $k = 0,...,K\text{-}1$ at which $f(\mathbf{x},t_k) < u_n$. We define the empirical distribution $F_N(y)$ as a step function with positive jumps of $1/N$ at the points $y_n= m(u_n)/K_A$, $n = 1, ..., N$, $F_N(0) = 0$.

The well-known hypothesis tests require that the function $F_N(y)$ be obtained by means of independent trials, i.e. that the random variables $u_n$, $n = 1,...,N$ be independent. This condition is not satisfied strictly, but there are reasons for assuming it to be approximately valid. Actually, large earthquakes are rare at high $M_0$ and the lattice cells in which they occur are widely spaced in time with a high probability. Two events in cells separated by a time interval $t$ are independent in the limit $t \to \infty$ because the seismic process possesses a decaying memory (the property of "mixing" in the terminology of random processes). The memory decay is ensured by such physical phenomena as stress relaxation due to the fracturing and viscosity of the medium and the healing of strength defects. When using a limiting property in the pre-limiting case one should keep in mind that the result will contain an uncontrollable error. However, the passage from statistical physics to thermodynamics uses a similar method known as the "thermodynamic limit" quite successfully. In this case a finite volume containing a finite number of particles is changed by infinitesimal volume with infinitely many particles. Therefore, we have reason to hope that the error at high $M_0$ can be neglected. An alternative is to refuse to test predictors at all. In this case, any predictor can claim to be suitable for prediction, and such arbitrariness is hardly acceptable in scientific research.

From this point of view, it is natural to examine only those predictors for which the hypothesis $H_0$ can be rejected at an acceptable level of significance. At large $N$, one can use the Kolmogorov statistic [Kolmogorov, 1933]

$$D_N = \sup |F(y) - F(y)|,$$

which has the asymptotic distribution

$$\lim_{N \to \infty} \Pr\{\sqrt{N} D_N < z\} = \sum_{k=-\infty}^{+\infty} (-1)^k e^{-2k^2 z^2}, \ z > 0$$

or the Smirnov statistic [Smirnov, 1938]

$$D_N^+ = \sup |F(y) - F(y)|,$$

$$D_N^- = \sup |F(y) - F(y)|,$$

with the asymptotic distributions

$$\lim_{N \to \infty} \Pr\{\sqrt{N} D_N^+ < z\} = \lim_{N \to \infty} \Pr\{\sqrt{N} D_N^- < z\} = 1 - e^{-z^2}, \ z > 0.$$

For small values of $N$, which is more usual in modern applications, the Smirnov statistics have the distribution [Smirnov, 1944]

$$\Pr\{D_N^+ \geq z\} = \Pr\{D_N^- \geq z\} = \sum_{k=0}^{[N(1-z)]} C_N^k z(z + \frac{k}{N})^{k-1} (1 - z - \frac{k}{N})^{N-k}, \ 0 < z < 1.$$

## 8. CONCLUSIONS

The two mathematical constructions introduced in this paper make it possible to develop a meaningful theory of prediction of large earthquakes. The first construction is the multicomponent random process with the properties of stationarity, ergodicity, locality, and conditional translational invariance (Section 3), which can take into consideration both temporal and spatial characteristics of seismicity. This in turn provides a solution to the prediction problem (Section 2), i.e., a method for estimating current and stationary probabilities of events by means of appropriate empirical rates (Section 4). The second construction is a generalization of the mathematical expectation, namely, the averaging not only over time (which, as a consequence of ergodicity, is the averaging over probabilities) but also over space. This method provides an opportunity to generalize to random fields such functions of two random variables as the correlation coefficient and the information on one random variable contained in another random variable and to obtain the corresponding characteristics of prediction efficiency (Sections 5 and 6). Another consequence of this generalization is the method of testing the predictive properties of a predictor (Section 7).

However, it is important to note that the accuracy of prediction, i.e., of the estimated probabilities of events, has substantial natural limitations. For short periods of time this accuracy depends on the number $m$ of large events that occurred during these periods in the region of prediction. An approximate estimator of the conditional probability of an event as a function of predictor values is a step function that has no more than $m + 1$ values. This is a consequence of small samples of large events. On the other hand, the nonstationary of natural seismicity over long time intervals in principle prevents the estimated conditional probability of an event from approaching its true value, so that an uncontrollable difference between these functions will inevitably be present.

An actual testing of the proposed method will be presented another paper of mine, which is under preparation for the publication in this journal.

## ACKNOWLEDGMENTS

I am grateful to I.V. Kuznetsov, S.A. Pirogov, V.F. Pisarenko, I.M. Rotvain, G.A. Sobolev, and M.G. Shnirman for helpful discussions.